\documentclass[aps,prl,twocolumn,superscriptaddress,showpacs,preprintnumbers,amsmath,amssymb,showkeys,nofootinbib]{revtex4-1}

\usepackage{amsmath}
\usepackage{amssymb}
\usepackage{soul}

\usepackage{enumerate}
\usepackage{graphicx}
\usepackage{dcolumn}
\usepackage{bm}
\usepackage{xcolor}
\usepackage{natbib,hyperref}

\begin{document}

\title{Beam Spin Asymmetry Measurements of Deeply Virtual $\pi^0$ Production with CLAS12}

\newcommand*{\ANL}{Argonne National Laboratory, Argonne, Illinois 60439}
\newcommand*{\ANLindex}{1}
\newcommand*{\CSUDH}{California State University, Dominguez Hills, Carson, CA 90747}
\newcommand*{\CSUDHindex}{2}
\newcommand*{\CANISIUS}{Canisius College, Buffalo, NY}
\newcommand*{\CANISIUSindex}{3}
\newcommand*{\SACLAY}{IRFU, CEA, Universit\'{e} Paris-Saclay, F-91191 Gif-sur-Yvette, France}
\newcommand*{\SACLAYindex}{4}
\newcommand*{\CNU}{Christopher Newport University, Newport News, Virginia 23606}
\newcommand*{\CNUindex}{5}
\newcommand*{\UCONN}{University of Connecticut, Storrs, Connecticut 06269}
\newcommand*{\UCONNindex}{6}
\newcommand*{\DUKE}{Duke University, Durham, North Carolina 27708-0305}
\newcommand*{\DUKEindex}{7}
\newcommand*{\FU}{Fairfield University, Fairfield CT 06824}
\newcommand*{\FUindex}{8}
\newcommand*{\FERRARAU}{Universita' di Ferrara , 44121 Ferrara, Italy}
\newcommand*{\FERRARAUindex}{9}
\newcommand*{\FIU}{Florida International University, Miami, Florida 33199}
\newcommand*{\FIUindex}{10}
\newcommand*{\FSU}{Florida State University, Tallahassee, Florida 32306}
\newcommand*{\FSUindex}{11}
\newcommand*{\GWUI}{The George Washington University, Washington, DC 20052}
\newcommand*{\GWUIindex}{12}
\newcommand*{\GSIFFN}{GSI Helmholtzzentrum fur Schwerionenforschung GmbH, D-64291 Darmstadt, Germany}
\newcommand*{\GSIFFNindex}{13}
\newcommand*{\INFNFE}{INFN, Sezione di Ferrara, 44100 Ferrara, Italy}
\newcommand*{\INFNFEindex}{14}
\newcommand*{\INFNFR}{INFN, Laboratori Nazionali di Frascati, 00044 Frascati, Italy}
\newcommand*{\INFNFRindex}{15}
\newcommand*{\INFNGE}{INFN, Sezione di Genova, 16146 Genova, Italy}
\newcommand*{\INFNGEindex}{16}
\newcommand*{\INFNRO}{INFN, Sezione di Roma Tor Vergata, 00133 Rome, Italy}
\newcommand*{\INFNROindex}{17}
\newcommand*{\INFNTUR}{INFN, Sezione di Torino, 10125 Torino, Italy}
\newcommand*{\INFNTURindex}{18}
\newcommand*{\INFNPAV}{INFN, Sezione di Pavia, 27100 Pavia, Italy}
\newcommand*{\INFNPAVindex}{19}
\newcommand*{\ORSAY}{Universit'{e} Paris-Saclay, CNRS/IN2P3, IJCLab, 91405 Orsay, France}
\newcommand*{\ORSAYindex}{20}
\newcommand*{\Juelich}{Institute fur Kernphysik (Juelich), Juelich, Germany}
\newcommand*{\Juelichindex}{21}
\newcommand*{\JMU}{James Madison University, Harrisonburg, Virginia 22807}
\newcommand*{\JMUindex}{22}
\newcommand*{\KNU}{Kyungpook National University, Daegu 41566, Republic of Korea}
\newcommand*{\KNUindex}{23}
\newcommand*{\LAMAR}{Lamar University, 4400 MLK Blvd, PO Box 10046, Beaumont, Texas 77710}
\newcommand*{\LAMARindex}{24}
\newcommand*{\MIT}{Massachusetts Institute of Technology, Cambridge, Massachusetts  02139-4307}
\newcommand*{\MITindex}{25}
\newcommand*{\MISS}{Mississippi State University, Mississippi State, MS 39762-5167}
\newcommand*{\MISSindex}{26}
\newcommand*{\ITEP}{National Research Centre Kurchatov Institute - ITEP, Moscow, 117259, Russia}
\newcommand*{\ITEPindex}{27}
\newcommand*{\UNH}{University of New Hampshire, Durham, New Hampshire 03824-3568}
\newcommand*{\UNHindex}{28}
\newcommand*{\NMSU}{New Mexico State University, PO Box 30001, Las Cruces, NM 88003, USA}
\newcommand*{\NMSUindex}{29}
\newcommand*{\NSU}{Norfolk State University, Norfolk, Virginia 23504}
\newcommand*{\NSUindex}{30}
\newcommand*{\OHIOU}{Ohio University, Athens, Ohio  45701}
\newcommand*{\OHIOUindex}{31}
\newcommand*{\ODU}{Old Dominion University, Norfolk, Virginia 23529}
\newcommand*{\ODUindex}{32}
\newcommand*{\JLUGiessen}{II Physikalisches Institut der Universitaet Giessen, 35392 Giessen, Germany}
\newcommand*{\JLUGiessenindex}{33}
\newcommand*{\URICH}{University of Richmond, Richmond, Virginia 23173}
\newcommand*{\URICHindex}{34}
\newcommand*{\ROMAII}{Universita' di Roma Tor Vergata, 00133 Rome Italy}
\newcommand*{\ROMAIIindex}{35}
\newcommand*{\MSU}{Skobeltsyn Institute of Nuclear Physics, Lomonosov Moscow State University, 119234 Moscow, Russia}
\newcommand*{\MSUindex}{36}
\newcommand*{\SCAROLINA}{University of South Carolina, Columbia, South Carolina 29208}
\newcommand*{\SCAROLINAindex}{37}
\newcommand*{\TEMPLE}{Temple University,  Philadelphia, PA 19122 }
\newcommand*{\TEMPLEindex}{38}
\newcommand*{\JLAB}{Thomas Jefferson National Accelerator Facility, Newport News, Virginia 23606}
\newcommand*{\JLABindex}{39}
\newcommand*{\UTFSM}{Universidad T\'{e}cnica Federico Santa Mar\'{i}a, Casilla 110-V Valpara\'{i}so, Chile}
\newcommand*{\UTFSMindex}{40}
\newcommand*{\INSUBRIA}{Universit\`{a} degli Studi dell'Insubria, 22100 Como, Italy}
\newcommand*{\INSUBRIAindex}{41}
\newcommand*{\BRESCIA}{Universit`{a} degli Studi di Brescia, 25123 Brescia, Italy}
\newcommand*{\BRESCIAindex}{42}
\newcommand*{\UCR}{University of California Riverside, 900 University Avenue, Riverside, CA 92521, USA}
\newcommand*{\UCRindex}{43}
\newcommand*{\GLASGOW}{University of Glasgow, Glasgow G12 8QQ, United Kingdom}
\newcommand*{\GLASGOWindex}{44}
\newcommand*{\YORK}{University of York, York YO10 5DD, United Kingdom}
\newcommand*{\YORKindex}{45}
\newcommand*{\VIRGINIA}{University of Virginia, Charlottesville, Virginia 22901}
\newcommand*{\VIRGINIAindex}{46}
\newcommand*{\WM}{College of William and Mary, Williamsburg, Virginia 23187-8795}
\newcommand*{\WMindex}{47}
\newcommand*{\YEREVAN}{Yerevan Physics Institute, 375036 Yerevan, Armenia}
\newcommand*{\YEREVANindex}{48}

\newcommand*{\NOWANL}{Argonne National Laboratory, Argonne, Illinois 60439}
\newcommand*{\NOWJLAB}{Thomas Jefferson National Accelerator Facility, Newport News, Virginia 23606}
\newcommand*{\NOWRUBochum}{Ruhr-Universit{\"a}t Bochum, 44801 Bochum, Germany}
\newcommand*{\NOWLIVERMORE}{Lawrence Livermore National Laboratory, Livermore, CA 94550}

\author {A. Kim} 
\affiliation{\UCONN}
\author {S. Diehl} 
\affiliation{\JLUGiessen}
\affiliation{\UCONN}
\author {K. Joo} 
\affiliation{\UCONN}
\author {V.~Kubarovsky} 
\affiliation{\JLAB}
\author {P.~Achenbach} 
\affiliation{\JLAB}
\author {Z.~Akbar} 
\affiliation{\VIRGINIA}
\affiliation{\FSU}
\author {J. S. Alvarado} 
\affiliation{\ORSAY}
\author {Whitney R. Armstrong} 
\affiliation{\ANL}
\author {H.~Atac} 
\affiliation{\TEMPLE}
\author {H.~Avakian}
\affiliation{\JLAB}
% ...

\collaboration{The CLAS Collaboration}
\noaffiliation

\begin{abstract}
	The new experimental measurements of beam spin asymmetry were performed for the deeply virtual exclusive $\pi^0$ production in a wide kinematic region with the photon virtualities $Q^2$ up to 8 GeV$^2$ and the Bjorken scaling variable $x_B$ in the valence regime.
	The data were collected by the CEBAF Large Acceptance Spectrometer (CLAS12) at Jefferson Lab with longitudinally polarized 10.6 GeV electrons scattered on an unpolarized liquid-hydrogen target.
	Sizable asymmetry values indicate a substantial contribution from transverse virtual photon amplitudes to the polarized structure functions.
	The interpretation of these measurements in terms of the Generalized Parton Distributions (GPDs) demonstrates their sensitivity to the chiral-odd GPD $\bar E_T$, which contains information on quark transverse spin densities in unpolarized and polarized nucleons and provides access to the proton's transverse anomalous magnetic moment.
	Additionally, the data were compared to a theoretical model based on a Regge formalism that was extended to the high photon virtualities.
\end{abstract}

\maketitle

\begin{figure}[b]
\begin{center}
	\includegraphics[width=0.85\linewidth]{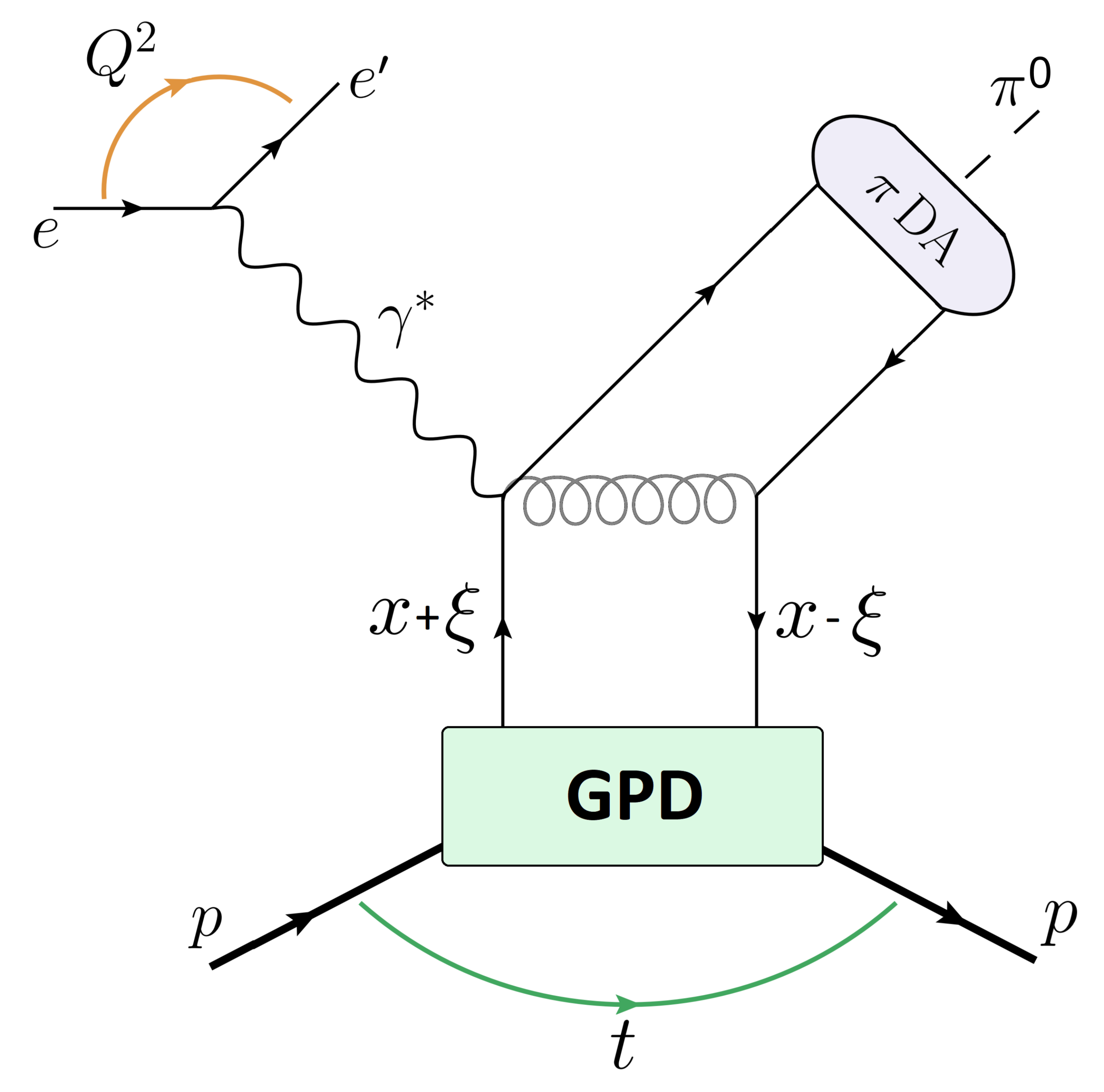}
	\caption{Hard exclusive electroproduction of a pion on the proton in very forward kinematics ($-t/Q^2 \ll 1$), described by GPDs~\cite{previous1, previous2}.}
	\label{fig:production_mechanism}
\end{center}
\end{figure}

Deeply virtual meson electroproduction (DVMP) is one of the most effective ways to access Generalized Parton Distributions (GPDs), which are essential non-perturbative objects that provide extensive information on the 3D structure of hadrons~\cite{Rad97-5, Col97-6, Brod94-7}.
DVMP processes at large photon virtuality can be factorized into a hard-scattering subprocess and a soft subprocess.
For longitudinally polarized virtual photons at large photon virtuality $Q^2$ the factorization of this amplitude shown in Fig.~\ref{fig:production_mechanism} has been proven~\cite{RADYUSHKIN1996333,Col97-6}.
For transversely polarized virtual photons, a modified perturbative approach is used in current phenomenological models to take the parton transverse momenta into account as a higher-twist effect~\cite{previous1}.
The hard subprocess can be calculated perturbatively and the soft parts of the convolution can be described with GPDs and a meson distribution amplitude (DA).

Previous experimental \cite{HERMES02, DeMasi2008, HERMES08, HERMES10, Bedlinskiy2012, Bedlinskiy2014, Kim2017, Bosted_pi02017, Bosted_piplus2017, Bedlinskiy2017, Zhao2019, hallA_2012, hallA_2016, hallA_2017, COMPASS20} and theoretical \cite{previous1, previous2, DK07, DMP17, SS19} studies of hard exclusive pseudoscalar meson electroproduction, especially $\pi^0$ and $\eta$ electroproduction~\cite{Bedlinskiy2014, Bedlinskiy2017, Kim2017, Zhao2019, previous1, previous2, Ahmad2009, Goldstein2011}, have shown that the asymptotic leading-twist approximation is not sufficient to describe the experimental results from the existing measurements.
It was found that there are strong contributions from transversely polarized virtual photons that have to be considered by including contributions from chiral-odd GPDs ($H_{T}$, $\widetilde{H}_{T}$, $E_{T}$, and $\widetilde{E}_{T}$) in addition to the chiral-even GPDs ($H$, $\widetilde{H}$, $E$, and $\widetilde{E}$), which depend on the momentum fraction of the parton $x$, the skewness $\xi$, and the four-momentum transfer to the nucleon $t$.
$\pi^0$ meson production was shown to have an increased sensitivity to chiral-odd GPDs and is especially suited to constrain $\bar E_{T}$, due to the quark flavor composition.

The chiral-even GPDs can be related to the well-known nucleon form factors \cite{previous2} but a few phenomenological constraints exist for the chiral-odd GPDs that cannot be accessed from the chiral-even sector.
For example, the first moment of $2 \widetilde{H}_{T} + E_{T}$ can be interpreted as the proton’s transverse anomalous magnetic moment \cite{Burk06}, and in the forward limit, $H_{T}$ becomes the transversity structure function $h_{1}$, which is directly related to the still unknown tensor charge of the nucleon \cite{previous2}.

An alternative description of hard exclusive pion production is provided by Laget (JML) model, which is based on Reggeized exchange of trajectories in the $t$-channel~\cite{JML20prog, JML20} and unitarity cuts \cite{JML10, JML11}.
While the Regge model starts at the real photon point and extends to the deeply virtual regime, a firm QCD foundation exists for the GPD model within the Bjorken regime and its applicability must be tested in the accessible $Q^{2}$ range.
For a precise comparison to theoretical models and especially for a study of higher-twist effects, a study in $t$, $\phi$, $x_{B}$, and $Q^{2}$ with multidimensional binning is needed to reduce uncertainties and to access the kinematic dependencies of the GPDs involved.

In exclusive meson production experiments, GPDs are typically accessed through differential cross sections and beam and target polarization asymmetries \cite{Dre1992, Are1997, Die2005}.
The focus of this work is on the extraction of the beam spin asymmetry moments related to the structure function ratio $\sigma_{LT'}/\sigma_{0}$.
In the one-photon exchange approximation the beam spin asymmetry (BSA) is defined as \cite{Dre1992, Are1997}:
\begin{eqnarray}\label{eq:BSA}
	BSA = \frac{\sqrt{2 \epsilon (1 - \epsilon)} \frac{\sigma_{LT^{\prime}}}
		{\sigma_{0}}\sin\phi}
	{1 + \sqrt{2 \epsilon (1 + \epsilon)}\frac{\sigma_{LT} }{\sigma_{0}} \cos\phi
		+ \epsilon \frac{\sigma_{TT}}{\sigma_{0}} \cos2\phi},
\end{eqnarray}
where the structure functions $\sigma_{L}$ and $\sigma_{T}$, which contribute to $\sigma_{0} = \sigma_{T} + \epsilon \sigma_{L}$, correspond to coupling to longitudinal and transverse virtual photons, and $\epsilon$ describes the flux ratio of longitudinally and transversely polarized virtual photons.
$\sigma_{LT}$, $\sigma_{TT}$, and the polarized structure function $\sigma_{LT^\prime}$ describe the interference between their amplitudes.
$\phi$ is the azimuthal angle between the electron scattering plane and the hadronic reaction plane in the center-of-mass frame.


For the present study, hard exclusive $\pi^0$ electroproduction was measured at Jefferson Lab with CLAS12 (CEBAF Large Acceptance Spectrometer for operation at 12 GeV) \cite{VDB20}.
Beam spin asymmetries in forward kinematics were extracted over a wide range in $Q^2$, $x_{B}$ and $\phi$.
The longitudinally polarized incident electron beam had an energy of 10.6~GeV with an average current of 40-55~nA, impinging on a 5-cm-long unpolarized liquid-hydrogen target placed at the center of the solenoid magnet of CLAS12.
The large acceptance of the CLAS12 detector allowed simultaneous detection of all four final state particles of the $\vec{e}p \rightarrow e^\prime p^\prime\pi^0$ reaction, with the $\pi^0$ reconstructed by measuring the 2$\gamma$ decay channel.
The scattered electron was identified in the forward detector using the track reconstructed in the drift chambers (DC) and matching it with signals in a lead-scintillator electromagnetic sampling calorimeter (EC) and Cherenkov counter.
The proton was identified as a positively charged particle track in the DC with the time-of-flight measurements from the scintillator counters.
The neutral pion decay photons were detected using the EC energy and timing information.

\begin{figure}[h!]
	\centering
	\includegraphics[width=0.95\linewidth]{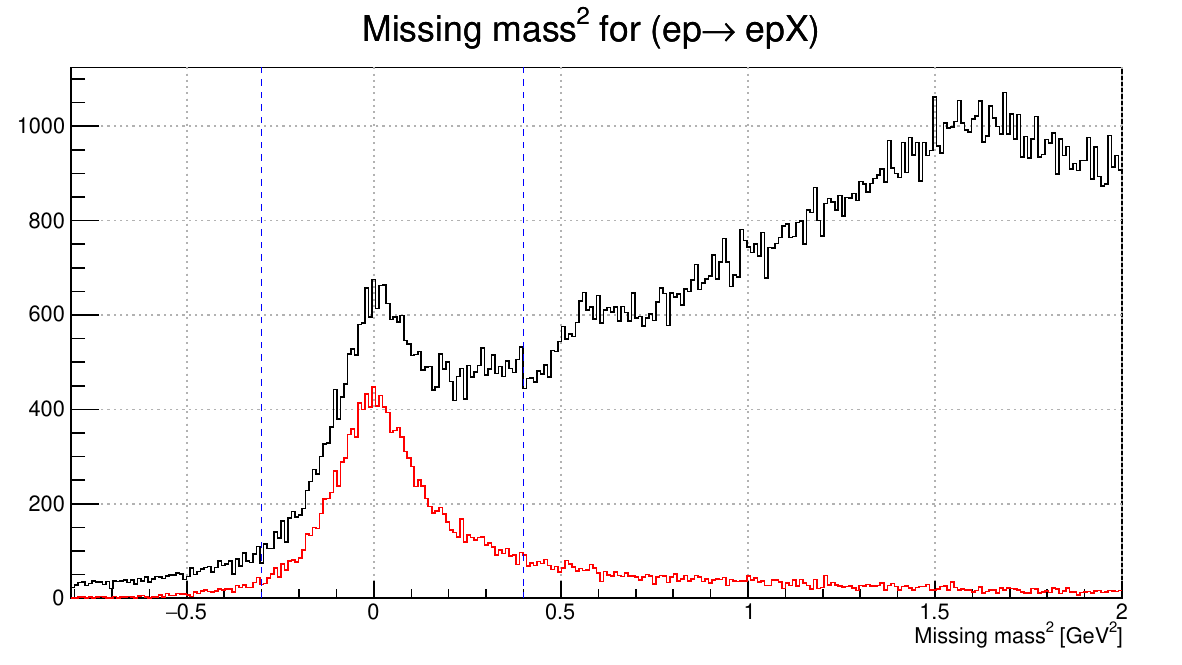}
	\caption{Distributions of missing mass squared of the $epX$ system before (black line) and after (red line) application of the exclusive constraints. The blue dashed lines represent the cuts on $MM^2_{epX}$ that were also used for final exclusive $\vec ep\to e^\prime p^\prime\pi^0$ event selection.
	}
	\label{fig:mm2beforeafter}
\end{figure}

For the selection of deeply inelastic scattered electrons, cuts on $Q^{2} > 2\, {\rm GeV}^{2}$ and on the invariant mass of the hadronic final state $W > 2$~GeV, were applied.
The events with exactly one electron, one proton and at least two photons were selected as candidates for the exclusive $\vec{e}p \rightarrow e^\prime p^\prime\pi^0$ final state.
With the 4-momenta reconstructed for all final state particles, the event kinematics is fully known, and energy and momentum conservation can be used to develop cuts to ensure exclusivity of the reconstructed events.
These constraints reject the backgrounds from different channels ({\it e.g.} $\eta$, $\rho$ or $\omega$ meson production) and from reactions with any additional undetected particle present.
The exclusivity cuts were based on the following variables:

\begin{itemize}
	\item $\left|\Delta P_T\right|<0.3$ GeV and $-0.5<\Delta P_z<0.9$ GeV - missing transverse and longitudinal momenta of the $e^\prime p^\prime\gamma\gamma$ system;
	\item $\left|\Delta\phi_{X\pi}\right|<4^\circ$ - the difference between the azimuthal angles of the reconstructed and computed $\pi^0$ using the beam, target, and reconstructed $e^\prime$ and $p^\prime$ particles, peaked around zero;
	\item $-0.3<MM^2_{epX}<0.4$ GeV$^2$ - missing mass squared of $epX$ system with the distribution peaked around the neutral pion mass squared.
\end{itemize}
The events within a $\pm3\sigma$ range from the expected peak values were chosen as the final exclusive candidates, where $\sigma$ is the observed experimental resolution obtained from the fit of each distribution.
Figure ~\ref{fig:mm2beforeafter} illustrates the effect of the $\Delta P_T$, $\Delta P_z$, and $\Delta\phi_{X\pi}$ cuts on the missing mass squared of the $epX$ system and demonstrates the power of these exclusive constraints to achieve clean $\vec ep\to e^\prime p^\prime\pi^0$ event selection.

\begin{figure}[h!]
	\centering
	\includegraphics[width=0.95\linewidth]{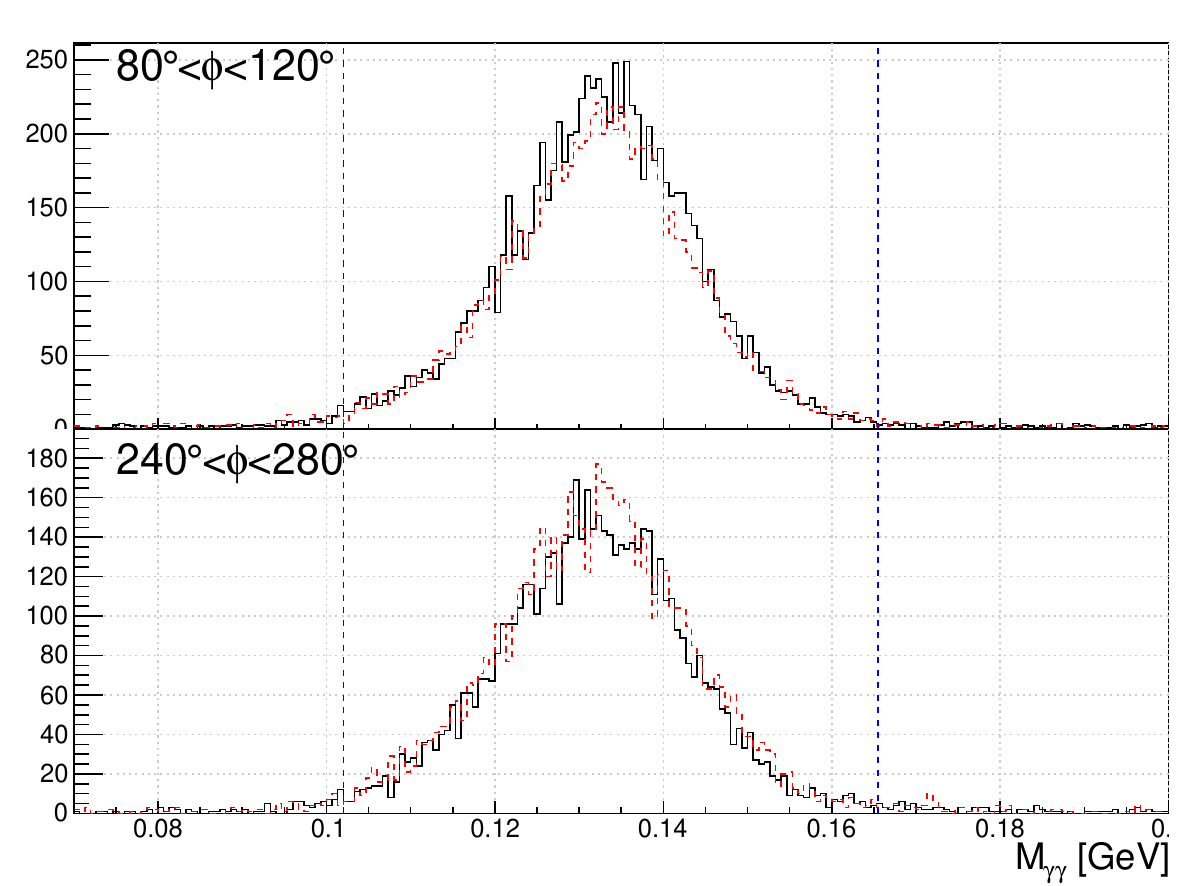}
	\caption{The invariant mass spectra of two decay photons show distributions peaked at the neutral pion mass.
The plots for two opposite $\phi$ bins are shown on top ($80^\circ<\phi<120^\circ$) and bottom ($240^\circ<\phi<280^\circ$).
The black solid histogram corresponds to the events with positive helicity and the red dashed histogram corresponds to the events with negative helicity.
The blue dashed lines represent $3\sigma$ cuts on the invariant mass of two photons.
The events outside of these lines are used to estimate the background for sideband subtraction.
}
	\label{fig:mgg}
\end{figure}

After application of all exclusivity cuts, the invariant mass of two photons was used to estimated the remaining background from accidental photons using the sideband method.
The observed background was found to be very small for all multidimensional bins, two of which are shown in Fig.~\ref{fig:mgg}.
As a cross-check, the $M_{\gamma\gamma}$ distributions were fit with a Gaussian (describing the signal) plus a first-order polynomial (describing the background).
The background estimate using the fit method was found to be consistent with the result from the sideband subtraction method, and was used to estimate the systematic uncertainty of the background subtraction.

\begin{figure}[b]
	\centering
	\includegraphics[width=0.95\linewidth]{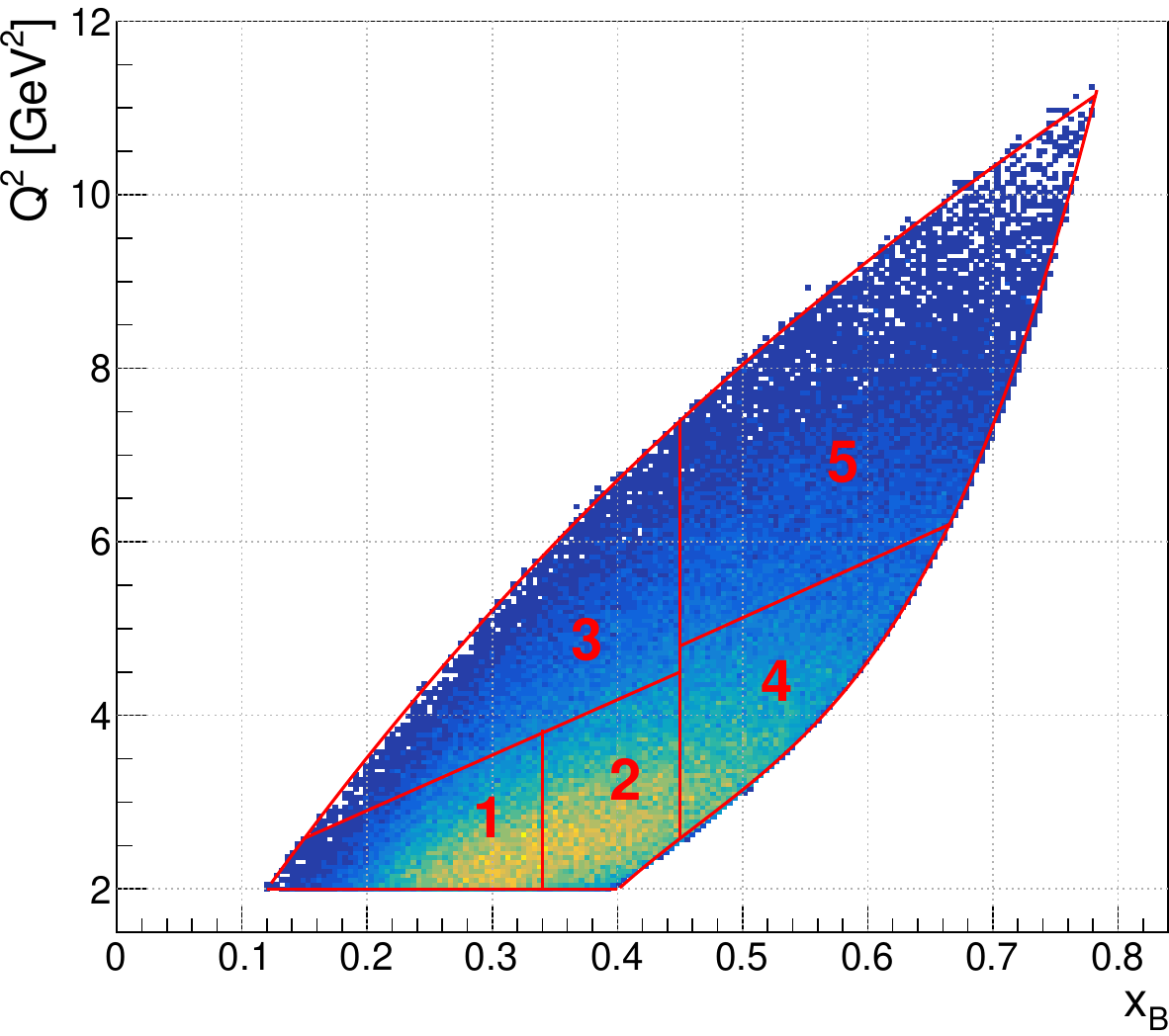}
	\caption{Distribution of $Q^{2}$ versus $x_{B}$. The red lines represent the bin boundaries, and the bin numbering is given.}
	\label{fig:q2x_bins}
\end{figure}

The BSA was determined experimentally from the number of signal counts with positive and negative helicity ($N^{\pm}_{i}$), in a specific bin $i$ as:
\begin{eqnarray}
	BSA_{i} = \frac{1}{P_{e}} \frac{N^{+}_{i} - N^{-}_{i}}{N^{+}_{i} + N^{-}_{i}},
\end{eqnarray}
\newline where $P_{e}$ is the average magnitude of the beam polarization.
$P_{e}$ was measured with a M{\o}ller polarimeter upstream of CLAS12 to be 86.3\%$\pm$2.6\%.
To obtain the signal counts, the $M_{\gamma\gamma}$ distribution for each multidimensional bin in $Q^{2}$, $x_{B}$, $-t$, and $\phi$ and for each helicity state was analyzed separately, and the background counts were subtracted using the sideband method, as described above.
Figure \ref{fig:q2x_bins} shows the $Q^{2}$ versus $x_{B}$ distribution of the exclusive events, together with the binning scheme applied for the multidimensional study.
The statistical uncertainty of the beam spin asymmetry was calculated based on standard error propagation.
For each of the five $\{Q^{2},x_{B}\}$ bins, three bins in $-t$ and nine bins in $\phi$ were defined to extract the BSA.

\begin{figure}[hbt]
	\centering
	\includegraphics[width=\linewidth]{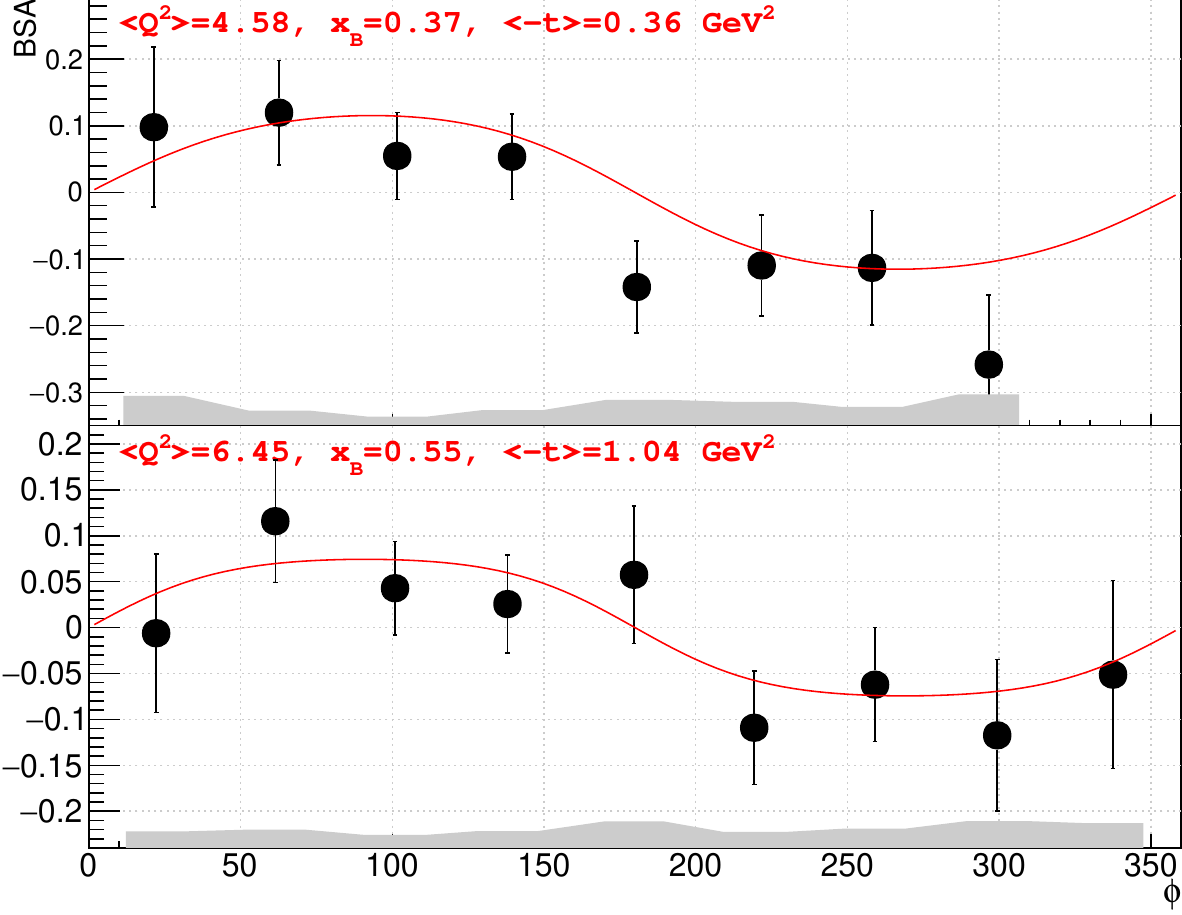}
	\caption{Beam spin asymmetry as a function of $\phi$ for two representative kinematic bins. The vertical error bars show the statistical uncertainty of each point. The gray bands represent systematic uncertainties of the BSA measurements. The red lines show the fit with  functional form of Eq.~(\ref{eq:BSA}).}
	\label{fig:sinphi_fit}
\end{figure}

To access the structure function ratio $\sigma_{LT'}/\sigma_{0}$, the BSA was plotted as a function of the azimuthal angle $\phi$.
Figure \ref{fig:sinphi_fit} shows the BSA as a function of $\phi$ in two exemplar $-t$ bins for two different $Q^{2}-x_{B}$ bin.
As expected, the $\phi$-dependence can be well described by Eq.~(\ref{eq:BSA}).
The denominator terms were fixed using the model parameterizations of the unpolarized structure functions measured by CLAS~\cite{Bedlinskiy2012}.
The impact of these terms in Eq.~(\ref{eq:BSA}) on $\sigma_{LT'}/\sigma_{0}$ was studied during the analysis using different parameterization values for the unpolarized structure functions and was found to be much smaller than the statistical uncertainty.

\begin{figure*}
	\centering
	\includegraphics[width=\linewidth]{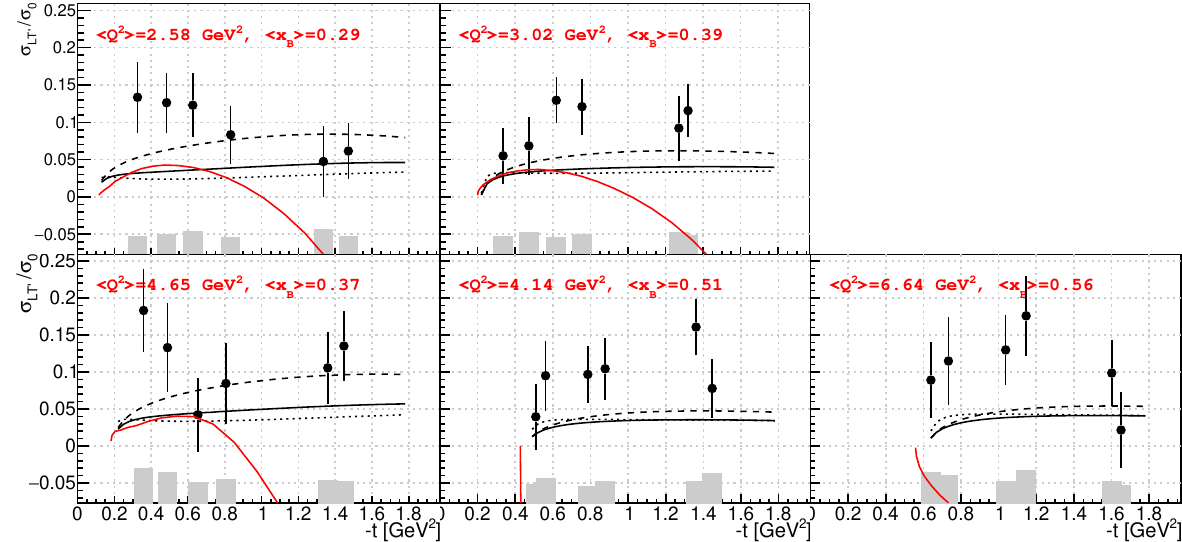}
	\caption{The measurements of $\sigma_{LT'}/\sigma_{0}$ and its statistical uncertainty as a function of $-t$ in the forward kinematic regime.
		The gray bins represent the systematic uncertainties.
	The black curves show the theoretical prediction from the GPD-based Goloskokov-Kroll model.
	The black dashed lines show the effect of the GPD $\bar E_{T}$ multiplied by a factor of 0.5, and the black dotted lines show the effect of the GPD $H_T$ multiplied by a factor 0.5.
     The red curve shows the theoretical predictions from the Regge-based JML model.}
	\label{fig:ALU_theory}
\end{figure*}

The extraction of the BSAs for the exclusive $\vec ep\to e^\prime p^\prime\pi^0$ channel includes several sources of systematic uncertainty.
Above we have discussed the contribution from the background subtraction, evaluated by using two different methods to estimate the background counts from the invariant mass distribution of the two decay photons.
The variations between asymmetries extracted using these two methods were 0.006 on average and were considered as systematic uncertainties.
The systematic effect due to the uncertainty of the beam polarization was determined to be around 0.003 based on the uncertainty of the measurement with the M{\o}ller polarimeter.
To estimate the impact of acceptance effects, a Geant4-based Monte Carlo simulation including CLAS12 detector effects was performed~\cite{GEANT4:2002zbu, UNGARO2020163422}.
The impact was evaluated by comparing the modeled and reconstructed asymmetries, and was found to be on the order of 0.013.
Also bin migration effects and radiative effects were studied based on Monte Carlo simulations and estimated to be around 0.002.
Additionally, for the systematic uncertainty associated with the event selection procedure, the exclusivity cuts were varied, and the corresponding BSA variations were estimated to be 0.014 on average.
As mentioned above, the effect of the denominator terms from Eq.~(\ref{eq:BSA}) on the fit results was also studied and estimated to be around 0.005.
The individual systematic uncertainties were combined in quadrature, and the total uncertainty is conservatively estimated at 0.015 on average, which is smaller than the statistical uncertainty in most kinematic bins.

Figure \ref{fig:ALU_theory} shows the final results for the BSA moments extracted in the region of $-t$ up to 1.6~GeV$^{2}$ for the five $\{Q^{2},x_B\}$ bins ($-t/Q^{2} \approx 0.2 - 0.4$), where the leading-twist GPD framework is applicable.
It includes the comparison to the theoretical predictions from the GPD-based model by Goloskokov and Kroll (GK)~\cite{GK09} and the Regge-based JML model~\cite{JML20prog, JML20}.
The structure function ratio $\sigma_{LT'}/\sigma_{0}$ is clearly positive in all kinematic bins and shaped by the contributing structure functions.
The non-$\phi$-dependent cross section $\sigma_{0} = \sigma_{T} + \epsilon \sigma_{L}$ is determined by the interplay between the $\bar E_T$ and $H_T$ contributions in the low $-t$ region, while $\sigma_{LT^\prime}$ is constrained to be zero at $-t_{min}$ due to angular momentum conservation.

The GK model includes chiral-odd GPDs to calculate the contributions from the transversely polarized virtual photon amplitudes, with their $t$-dependence incorporated from Regge phenomenology.
The GPDs are constructed from double distributions and constrained by the latest results from lattice QCD and transversity parton distribution functions \cite{GK09}.
A special emphasis is given to the GPDs $H_{T}$ and $\bar{E}_{T} = 2 \widetilde{H}_T + E_T$, while contributions from other chiral-odd GPDs are neglected in the calculations, unlike chiral-even GPDs.
$\sigma_{LT^\prime}$ can be expressed through the convolutions of GPDs with subprocess amplitudes (twist-2 for the longitudinal and twist-3 for the transverse amplitudes) and contains the products of chiral-odd and chiral-even terms \cite{previous1}:
\begin{eqnarray}
	\sigma_{LT^\prime} \sim \xi \sqrt{1-\xi^{2}} \frac{\sqrt{-t'}}{2m} Im[ \langle\bar{E}_{T}\rangle^{*} \langle\widetilde{H}\rangle + \langle H_{T}\rangle^{*} \langle\widetilde{E}\rangle].
	\label{eqn:sigma_GPD}
\end{eqnarray}
After expanding the dominating chiral-odd denominator term~\cite{previous1}, the structure function ratio $\sigma_{LT^\prime}/\sigma_0$ can be expressed by:
\begin{eqnarray}
	\frac{\sigma_{LT^\prime}}{\sigma_0} \sim\frac{Im[ \langle\bar{E}_{T}\rangle^{*} \langle\widetilde{H}\rangle + \langle H_{T}\rangle^{*} \langle\widetilde{E}\rangle]}
	{(1- \xi^{2}) \left|\langle H_{T} \rangle\right|^{2} - \frac{t'}{8m^{2}} \left|\langle \bar{E}_{T} \rangle\right|^{2} + \epsilon\sigma_L}.
\end{eqnarray}
Due to the quark flavor composition of the pions, $\pi^{0}$ production is typically dominated by $\bar E_{T}$, while the contribution from ${H}_{T}$ is significantly smaller.
In contrast, $\pi^{+}$ electroproduction shows a significantly stronger contribution from ${H}_{T}$.
Since chiral even GPDs are much better known than their chiral odd counterparts, the strongest uncertainty for the theoretical prediction is expected from the so far poorly known GPD $\bar E_{T}$.

The comparisons between the experimental results and theoretical calculations demonstrate the difficulty to parameterize the delicate interference structure function $\sigma_{LT^\prime}$ and estimate its sizable magnitude.
The JML model shows positive values for the beam spin asymmetries in the three lowest $x_B$ (close to 0.35) and $Q^2$ (below 4.5 GeV$^2$) bins for the low $-t$ regions, but fails to extrapolate to the two highest $x_B$ and $Q^2$ bins.
The GK model provides a better description of the experimental measurements in a wide $Q^2$ and $-t$ range, but still predicts significantly smaller values for $\sigma_{LT^\prime}/\sigma_0$.
This discrepancy between the GK predictions and the experimental data might be explained by the interplay between the magnitudes of the chiral-odd GPDs $H_{T}$ and $\bar E_T$.
Based on Eq.~(\ref{eqn:sigma_GPD}) the results especially hint that $\bar E_{T}$ is overestimated.
To illustrate the sensitivity of $\sigma_{LT'}/\sigma_{0}$ on the GPD $\bar E_{T}$, Fig.~\ref{fig:ALU_theory} also contains calculations with the GPD $\bar E_{T}$ reduced by an overall factor of 2 (black dashed line) and with the GPD $H_T$ reduced by a factor 2 (black dotted line).
The modification of the GPD $\bar E_{T}$ generates substantially larger BSA values, whereas the reduction of the GPD $H_T$ shows a significantly smaller effect.
This disparity reflects the dominance of the GPD $\bar E_T$ in the theoretical description of $\pi^0$ electroproduction, which makes it the most relevant channel to constrain $\bar E_T$.
These effects are especially evident for the lower $Q^2$ bins, while the increase in the high $Q^2$ bins is noticeably smaller, which can indicate that the contributions of chiral-odd GPDs are still significant at the range of $Q^2$ accessible in CLAS12, and should be improved in the GK model calculations.

While a change of $\bar E_{T}$ helps as far as the description of $\sigma_{LT'}/\sigma_{0}$ is concerned, the consequences for other observables remain to be checked.
This includes the measurements that show strong contributions from the transversity GPDs and need to be considered for the determination of $\bar E_{T}$, such as unpolarized cross section measurements for deeply virtual $\pi^0$ production from CLAS~\cite{DeMasi2008, Zhao2019, Bedlinskiy2012, Bedlinskiy2014, Bedlinskiy2017}, Hall A~\cite{hallA_2012, hallA_2016, hallA_2017}, COMPASS~\cite{COMPASS20}, and observables with transversely polarized targets for hard exclusive $\pi^{+}$ production from HERMES~\cite{GK09}.
Altogether, a new global fit of the GPDs to all existing data from CLAS and Hall A, as well as the aforementioned HERMES and COMPASS results, and additional upcoming CLAS12 results on other mesons, becomes necessary.
Here, the new multidimensional precision $\pi^{0}$ BSA data from this work and its high sensitivity to the GPD $\bar E_{T}$ will allow a better determination of this so far poorly known GPD.
Based on the improvements in the knowledge of $\bar E_{T}$, it will become possible to improve the knowledge of the nucleon's anomalous magnetic moment $k_{T}^{u,d} = \int dx \bar E^{u,d}_{T}(x, \xi, t=0)$, which is a fundamental quantity and so far only poorly constrained using lattice QCD results.


In summary, we have performed a multidimensional study of the BSA measurements for $\vec{e}p \to e^\prime p^\prime \pi^0$ at large photon virtuality, above the resonance region.
In very forward kinematics, the magnitude of $\sigma_{LT'}/\sigma_{0}$ is underestimated in all $Q^{2}$ and $x_{B}$ bins by the most advanced GPD-based model \cite{GK09}, indicating that a global fit of the model to existing experimental data is necessary to achieve an improved parameterization of the chiral odd GPDs, especially the dominating GPD $\bar E_T$.

We acknowledge the outstanding efforts of the staff of the Accelerator and the Physics Divisions at Jefferson Lab in making this experiment possible.
We owe much gratitude to P. Kroll for many fruitful discussions concerning the interpretation of our results.
This work was supported in part by the U.S. Department of Energy, the National Science Foundation (NSF), the Italian Istituto Nazionale di Fisica Nucleare (INFN), the French Centre National de la Recherche Scientifique (CNRS), the French Commissariat pour l'Energie Atomique, the UK Science and Technology Facilities Council, the National Research Foundation (NRF) of Korea through grants provided by the Ministry of Science and ICT, the Helmholtz-Forschungsakademie Hessen f{\"u}r FAIR (HFHF), the Deutsche Forschungsgemeinschaft (DFG), and the Chilean Agency of Research and Development (ANID).
The Southeastern Universities Research Association (SURA) operates the Thomas Jefferson National Accelerator Facility for the U.S. Department of Energy under Contract No. DE-AC05-06OR23177.

\bibliographystyle{apsrev4-1}
\bibliography{main}


\end{document}